 \documentstyle[12pt,aps,floats]{revtex}
\begin{document}
\font\mine=cmex10
\def\check{\raise12pt\hbox{\mine p}}

\title{Prioritizing Science: A Story of The Decade Survey for the
1990s\footnote{To be published in the Centennial Volume of the American
Astronomical Society, David DeVorkin editor (American Institute of
Physics).
  This article is based
upon a shorter report published in Science, {\bf 251}, 1412, 1991. }}

\author{John N. Bahcall}

\maketitle

\bigskip
What are the most important aspects of the universe to explore? 
What are the best ways to make
discoveries in astronomy and astrophysics? These are
tough questions because researchers have many different approaches
and it is usually not clear, until the most interesting problems are
solved, which method will yield the most important results. 
Individual astronomers present strong arguments for many potential
approaches that require federal funding.

We are well into an era of limited research budgets, however, and
choices have to be made. We astronomers have recognized that if we 
do not set our own priorities, then funding agencies and
congressional 
officials will do it for us, and will do it less well. 
Moreover, we have learned over the years that the process of
 trying to convince colleagues in different specialties both 
improves our projects 
and provides a broader and more reliable basis for support.

The Decade Survey of Astronomy and Astrophysics 
for the 1990s gave specific answers to the hard
questions of what to fund and, by implication, what to cut. Working
under the auspices of the National Research Council, 
we astronomers---acting as a community---recommended funding 
for a limited number of initiatives,
ranked in order of priority. Only one out of every ten highly
promising initiatives survived this rigorous selection.

We have been spectacularly successful in getting a very large fraction
of our recommendations implemented, as can be seen from
the tables that are collected near the end of this article. The tables summarize
the fate of the projects we ranked highly.

In this article, I will describe, from my perspective as chairman of
the survey committee for the 1990s, how we came to a consensus on what
to recommend. I
hope that an understanding of our experience may be useful in future surveys.

\section{The Survey Committee}

The group charged with setting priorities, the Astronomy and
Astrophysics Survey Committee, was established by
the National Research Council (NRC) in May 1989, following my
appointment as chair in February 1989.  The membership of the
committee is shown in Table~\ref{comm}. The report of the committee, 
{\it The Decade of Discovery in Astronomy and Astrophysics}, was
published in March 1991 by the National Academy Press.

The first step was to find an outstanding group of scientists who
were willing to sacrifice a significant part of their research time in
order to serve on the committee. I spent most of the months
between February and May of 1989 talking to hundreds of astronomers 
about potential members who might serve on the advisory
panels of the survey and on the executive committee (hereafter, the
survey committee). I also wrote to the chair of every astronomy
department in the United States, as well as to many other prominent
astronomers, requesting nominations. I invited each person to
suggest themes and questions for the study. In addition, I wrote to
a number of distinguished astronomers abroad asking about astronomical
programs in their countries and requesting advice about
possible international collaborations.

\begin{table}[htb]
\centering
\tighten
\begin{minipage}[htb]{5.25in}
\caption[]{Astronomy and Astrophysics Survey Committee\protect\label{comm}}
\begin{tabular}{ll}
John N. Bahcall&Institute for Advanced Study, {\it Chair}\cr
Charles A. Beichman&Institute for Advanced Study, {\it Executive
Secretary}\cr
Claude Canizares&Massachusetts Institute of Technology\cr
James Cronin&University of Chicago\cr
David Heeschen&National Radio Astronomy Observatory\cr
James Houck&Cornell University\cr
Donald Hunten&University of Arizona\cr
Christopher F. McKee&University of California, Berkeley\cr
Robert Noyes&Harvard-Smithsonian Center for Astrophysics\cr
Jeremiah P. Ostriker&Princeton University Observatory\cr
William Press&Harvard-Smithsonian Center for Astrophysics\cr
Wallace L. W. Sargent&California Institute of Technology\cr
Blair Savage&University of Wisconsin\cr
Robert W. Wilson&AT\&T Bell Laboratories\cr
Sidney Wolff&National Optical Astronomy Observatories\cr
\end{tabular}
\end{minipage}
\end{table}

I filled nearly all of a looseleaf notebook with comments made by
astronomers about the judgment and vision of their colleagues.  (I
later put this notebook into a secure trash can out of respect for the
confidential nature of these conversations.)  After many
conversations, consistent pictures emerged. 
Some astronomers and astrophysicists were always mentioned with the  
highest respect by the people with
whom they worked.  These were the scientists we wanted to serve and,
with only two exceptions out of our first 15 choices, nearly everyone
I asked agreed to make the required personal sacrifice.

I am convinced that the unanimity which was achieved among astronomers
in support of our recommendations was due in large part to the
scientific distinction and judgment of my colleagues on the
committee.  In retrospect, I believe I could have assembled a
different committee of 15 people with approximately equal scientific
qualifications.  But, I am certain that I could not have improved
significantly upon the high level of respect of the 
astronomical community for the achievements and
insights of the committee members.  This respect was our most
important asset.

The 15 members of the survey committee were nominated by the
appropriate committees of the National Research Council and were
appointed by Frank Press, the chairman of the National Research
Council and the president of the National Academy of Sciences. 
The survey committee contained six members of the
National Academy of Sciences, two Nobel Prize winners, and two
directors of national observatories. 
In addition to the committee itself, Frank Press  
took an active role in supporting and representing the work of the
decade committee within Washington.
(For astronomers not familiar with Frank's past, I note that
he is a distinguished geophysicist and and a former
national science advisor to President Carter. Throughout the survey,
Frank provided immediate access, strong support, and valuable 
guidance and insights.)

I believe that the most important decision we made as a survey
committee was to base our recommendations on scientific merit,
independent of political considerations.  We had credibility as
scientists judging science.  We did not have special expertise in
guessing which way the political winds would blow.  Early in the
survey process, some individuals who worked for a federal 
agency tried to tutor us in the best
political strategies, but I do not remember ever a single case in
which these tutorials were repeated after we made clear that we were
limiting the criteria for our prioritization to scientific merit.  I
believe that our insistence in judging only science greatly simplified
our task and contributed to the lasting value of the survey report.

If I could made just one recommendation to future participants in
decade surveys it would be:  ``Stick to the science.''  Other
committees and studies are charged with the responsibility for
devising the best possible strategy for implementing in a timely way
the science priorities.  Only the decade committee is charged with
setting the science goals for that decade.

\section{Washington Visits}

Prior to the formation of the survey committee, Frank Press and
I visited major agency heads and congressional and administration
leaders in order to obtain their advice on what issues the report
should address and in what form the results should be presented. I
did not ask for support of any projects on these ``get-acquainted''
visits, but I did hope to create a
favorable climate for future consideration of astronomy initiatives. I
also did not ask what answers would be politically most desirable.
Participants in the survey were encouraged to solicit facts from
agency and administration authorities, but we evaluated ideas and
initiatives independently and in confidence. Agency leaders,
congressional 
staffers, senior people at the Office of Management and
Budget, and the President's science advisor (who had gone through
a similar experience as chair of a previous NRC decade survey for
physics) all provided valuable advice.

The consultations in Washington ultimately resulted in several important
sections of the final report: a chapter on the lunar initiative
(requested by the Administrator of NASA); a
chapter on high speed computing (suggested by the Director of NSF); 
an emphasis on priorities for
technology in this decade that will lead to science in the next decade
(proposed by the Deputy Administrator for Space Science of NASA);
recommendations of what astronomers should do {\it pro bono} to help
with the crisis in education (requested by Dick Mallow, then a senior 
Congressional
staffer and now a senior AURA officer);  an examination of the technical heritage of proposed 
initiatives (requested by people at OMB); realistic estimates of the costs for each of the new 
projects; a chapter on astronomy as a national
asset; an
examination of the role of American astronomy in the international
context with some guidelines for assessing when international 
collaborations would be fruitful; and 
thumb-nail sketches of major projects
that could be used conveniently by staffers helping to draft legislation.
Where I have not made specific attributions, similar suggestions were
made by several different people we visited.

\section{The Panels}

The first task of the survey committee was to select  the
chairs of 15 advisory panels for different subdisciplines, based on
discussions with astronomers of different specialties at institutions
throughout the country. 
The survey committee 
decided that  the subject matter covered by the different panels
should reflect the subdivisions that
astronomers generally use in identifying their specialties, especially
 the wavelength or technique used to make astronomical observations. 
It would have been more logical to have organized the panels along
scientific goals, independent of wavelength or 
technique, but I am convinced that
the more logical organization would have been less effective. The
similar viewpoints and experiences 
that were shared by people within a given technical subdivision of
astronomy made it easier to reach a scientific consensus. The survey
committee was responsible for integrating the advice by the different
discipline panels and formulating the highest priority science program.
 
Future decade surveys may well choose to organize themselves
differently.  
The revolutionary opportunities provided by, for example, the 
VLA, the HST, COBE, x-ray satellites, and the Keck telescopes have all
forced astronomers to realize that in order to solve scientific
problems they have to use different techniques.  Many of the younger
astronomers now identify themselves in terms of the scientific problems
they work on rather than the techniques they use to solve problems. I
think this is a healthy development.

Table~\ref{panel} lists the  chairs of our panels.  In choosing
the panel chairs, the survey committee again used scientific
distinction and widely respected judgment as the primary selection
criteria, but we also took account of the necessity of obtaining
expert advice about the major research fields and techniques.

The panel chairs and the survey committee jointly 
selected 300 people for the advisory groups. The members of these
groups had a high level of
scientific achievement and also represented different research
approaches, different kinds of institutions, and different geographical areas.
The survey committee
itself considered projects that spanned more than one subfield or
which fell between the assigned responsibilities of the panels.

The panels met at different sites in the United States in order to
help stimulate wide participation by the astronomical community. I
also wrote to each of the panel members and asked them to solicit the
views of colleagues at their home institutions. Local discussions of
issues in individual astronomy and physics departments generated
valuable ideas and helped consolidate support for the 
final recommendations of the decade survey.

\begin{table}[t]
\centering
\tighten
\begin{minipage}[htb]{6.5in}
\caption[]{Astronomy and Astrophysics Survey Panels\protect\label{panel}}
\begin{tabular}{ll}
\multicolumn{1}{c}{Panel}&\multicolumn{1}{c}{Chair}\cr
\noalign{\medskip\hrule\medskip}
Benefits to the Nation&Virginia
Trimble\cr
\ \ \ \ \ \  from Astronomy and Astrophysics\cr
Computing and Data Processing&Larry Smarr\cr
High Energy from Space&Bruce Margon\cr
Infrared Astronomy&Frederick Gillett\cr
Interferometry&Stephen Ridgway\cr
Optical/IR from Ground&Stephen Strom\cr
Particle Astrophysics&Bernard Sadoulet\cr
Planetary Astronomy&David Morrison\cr
Policy Opportunities&Richard McCray\cr
Radio Astronomy&Kenneth I. Kellermann\cr
Science Opportunities&Alan Lightman\cr
Solar Astronomy&Robert Rosner\cr
Status of the Profession&Peter B. Boyce\cr
Theory and Laboratory Astrophysics&David N. Schramm\cr
UV-Optical from Space&Garth Illingworth\cr
Working Group on Astronomy from the Moon&Charles A. Beichman and
John. N. Bahcall\cr
\end{tabular}
\end{minipage}
\end{table}

Much of the difficult work of the survey was done within the panels.
The panel chairs (see Table~\ref{panel}) were forced to exercise tact, 
scientific insight, and organizational skills. 
Scientists from groups that had
traditionally worked in competition with each other had to develop
unified recommendations.  Perhaps the most
difficult task, brilliantly achieved, 
was to form a consensus set of recommendations within
the optical and infrared community (Chair: Steve Strom).  The lack of consensus in this
community had frustrated attempts to obtain some needed 
major facilities in preceding surveys.

The panels began their technical work with essays submitted by individual 
panel members on
what they identified as the most important issues or projects.
After the essays were discussed, 
there were  presentations by panel members or by invited outside
experts on all the significant questions that were to be included 
within the text of the panel's report.  A core group within each panel
wrote the initial draft of the report, which was then iterated within
the panel and then commented on by members of the survey committee.

The most intense discussions in the first nine months of the survey
occurred within the panels. In order to ensure good communication
between the panels and the survey committee, each member of the
survey committee served as the vice-chair of one of the panels. This
arrangement worked well, keeping the survey committee apprised of
ideas as they developed and enabling each panel to understand the
goals and procedures of the full survey.

Because of their special responsibilities, two of the panels operated 
differently from the others.
The beautiful non-technical chapters on Science Opportunities and on 
Astronomy as a National Asset
were 
written almost entirely by the relevant panel chairs  (Alan Lightman and
Virginia Trimble, see Table~\ref{panel}), with advice and comments
from their panel members.
We were fortunate to have in charge of these activities
 accomplished research astronomers who are also 
outstanding writers.

The survey committee avoided many potential problems by
deciding that the panel reports would be advisory rather than part of
the findings of the survey and that the reports would not be refereed
by either the survey committee or by the NRC. The 
recommendations of the panels were not 
binding on the survey committee, but
the panel reports contain important technical information, as well as
detailed arguments advocating specific initiatives. The reports of the
panels were published separately from, but simultaneously with, the
full survey report by the National Academy Press under the title
{\it Working Papers: Astronomy and Astrophysics Panel Reports}.

\section{Developing a Consensus}

I believe that it was essential, in forming a consensus,  
to involve the community as much as
possible.
The survey was organized so that 
every astronomer who had something to say had an
opportunity to be heard. 

Open discussions were held in conjunction
with meetings of the American Astronomical Society (AAS) and at
several other professional society meetings. In January 1990, at the 
Washington, D.C., meeting of the AAS, nearly 1000 astronomers 
participated in open sessions that involved all 
15 of the panels. The names
of the survey committee members and of the chairs of the panels
were published in the AAS newsletter, along with remarks encouraging
individual astronomers to present their ideas directly to survey
committee members, panel chairs, or panel members.

Establishing the recommendations of the survey took 14 months,
about a year less than was projected. 
The committee worked efficiently because we were  busy scientists 
eager to get back to our research, because we had effective leadership
and support from the National Research Council (Robert L. Riemer) and from
the Executive Secretary of the survey (Chas Beichmann), and because
we were the first decade survey to be able to rely on  
fax and email communications for many of our discussions and iterations
of texts.

The survey committee had six
meetings at astronomical centers throughout the country.

I was surprised by one thing. Veterans of similar activities assured
me that there would be a difficult and tense period of bargaining
before we agreed on the final recommendations. This never 
happened. I believe the reason is that the committee judged the initiatives on the basis of scientific potential,
 without regard to political considerations.

The list of priorities was established by a gradual process that was
much easier than any of us anticipated.
The committee voted on straw ballots on three occasions during our regularly 
scheduled meetings, using as
background material the preliminary reports of the advisory panels.
The straw ballots focused the discussion on projects that were most
likely to be considered important in the final deliberations. As 
preparation for the final ballot, the committee heard advocacy
presentations from the panel chairs. The chairs also participated in
discussions of the relative merits of all the initiatives, although the
final recommendations were formulated by the survey committee in
executive session.

Two strategic decisions helped the committee reach a consensus
quickly and smoothly. First, the committee decided that if we failed
to reach agreement in July 1990 at the pleasant facilities of the
National Academy, within reach of cool Pacific breezes 
in Irvine, California, then we would meet a month later in the least
desirable place in the middle of summer that we could think of,
namely, Washington, D.C.

Second, several committee members proposed that I draw up, on
the evening before the final voting, a draft list of recommended
initiatives in order of priority. They suggested that the committee
alter by consensus the draft set of recommendations in order to
arrive at the final list of priorities. The proposers hoped that, by this
process, the committee could avoid having ``winners or losers." I was
skeptical of the chances for success when the idea was proposed, but
I agreed to try.

Having drawn up a handwritten list of priorities on the night
before our formal voting, I was surprised the next day at how rapidly
we reached a consensus. We began with those equipment categories
concerning which we were most in agreement and then worked our
way to the more difficult choices. We went around the table,
everyone stating their views about what change, if any, needed to be
made in the ordered list that we were considering. By the time we
had all spoken, the consensus was obvious and we adopted 
unanimously our priorities in each category.

In times of budgetary crisis, good citizenship also requires fiscal
restraint. The survey committee studied approximately ten times as
many initiatives as were endorsed, recommending that funding
agencies invest in astronomical initiatives according to the scientific
priorities established in the survey report.

\section{The Survey Report}

The 180 page book presenting the recommendations was written
in about three months. National Research Council reports are
reviewed carefully. They must meet high standards of logic, of
evidence, and of objectivity. In our case, the National Research
Council selected 18 formal referees, in addition to a report review
committee. The reviewers were anonymous National Academy
members and other qualified scientists, in physics, in astronomy, and
in other related disciplines. The formal review process was painful,
but I answered each review comment, even rhetorical questions,
with a specific written response in order that we could complete the
review quickly. The 18 referees helped to sharpen our arguments
and to clarify our logic, but did not suggest revisions of our
priorities.

Ours was the fourth in a series of decade surveys by astronomers, led
by A. Whitford, J. Greenstein, and G. Field, respectively. The
highest priority initiatives in each survey were successfully 
undertaken, encouraging astronomers to submerge parochial interests and
focus on the most important initiatives.

Would another committee of astronomical experts have 
recommended a similar set of priorities? 
I think so, provided that they had
also spent a year learning about and comparing all the proposed
initiatives in this country and abroad.

The report was published under the title {\it The Decade of Discovery in
Astronomy and Astrophysics} by the National Academy Press in
Washington, D.C. in 1991.  It is still available, and I believe still
good reading.

\section{Categories of Recommendations}

How many categories of recommendations should we have?  Should we
make separate recommendations for space missions and for ground based
projects? Should we have recommendations that prioritized projects 
independently of the potential funding agency ?

These were the most hotly debated issues we faced in the survey.  

In preliminary discussions, most agency personnel opposed 
absolute rankings that combined ground and space initiatives, worrying
that their top priorities might be adversely affected by ineffectiveness
at some other agency. 
We decided not to yield to these worries.

We decided to provide an overall 
prioritization independent of agency or of technique, because
we believed  that good citizenship required us to use our scientific
expertise to provide the maximum possible guidance to those
responsible for making budgetary decisions.
We also 
provided separate recommendations within the categories of Space
projects and Ground Based projects.
In addition, we  decided on a common sense division of
recommendations into Large, Medium, and Small, based upon the
financial resources required to achieve the projects.

I believe that our decision to provide an overall prioritization was
correct and increased the credibility of the survey report in
Congress and in the executive branch, particularly at the Office of
Management and Budget.  Some high ranking agency officials predicted
dire consequences for programs under their responsibility if we
insisted on prioritizing across the space (NASA) and ground based
(NSF, DOE) categories. They feared that their favorite 
programs would be held
hostage to ineffectiveness or budgetary constraints 
at other agencies. 
As far as I know, 
the predicted difficulties did not occur.
Recommended programs were not delayed inappropriately 
because of the competition 
for higher prioritization between ground based and
space based projects.

\section{Our Recommendations and How They Fared}

What did we recommend? What was achieved?

The committee assigned its highest priority for ground-based
astronomy to the revitalization of the infrastructure for research,
both equipment and people. 

It is difficult to assess quantitatively the effect of this 
recommendation for infrastructure support.  
Unlike new projects which are either funded or not 
(see Tables~\ref{lgpro}--\ref{smallpro} below), the support actually achieved for
infrastructure has to be judged against what that support would have
been in the absence of our recommendation.  We can never really know
how bad the situation would have been in the absence of a strong
statement by the survey committee. Our strong endorsement of
infrastructure support has often been cited in discussions within NSF
committees and with congressional staffers.  Enthusiastic and
energetic senior staff members at NSF and NASA, as well as our
colleagues on the  NRC Astronomy and
Astrophysics Committee, have repeatedly used the high priority assigned
to infrastructure support in the Decade Survey to argue for the maximum
possible resources being directed to astronomy for individual grants
and for maintenance of existing equipment.

My own assessment is that our recommendation for infrastructure
support for people and existing equipment was wonderfully successful
with NASA (which has become the principal supporter of astronomical
sciences in the 1990s) and only modestly successful at NSF. 
The freedom to redirect resources at NSF was limited in part by the large
capital investments for LIGO (gravity wave detector) and the Gemini 
(northern and southern 8-m telescopes) programs.

Continuing to develop a space program
with an improved balance between large and small projects, with
emphasis on quicker and more efficient missions, was the 
committee's highest priority for space research.
This recommendation resonated with the views of the current NASA
administrator (Dan Goldin) and has become a theme of the NASA
astrophysics program.

I now want to review informally the fate of our priority
recommendations for individual projects.  

We recommended four large programs, which are shown in Table~\ref{lgpro}.  All
four of these programs are being developed!
In addition, the AXAF
observatory, recommended by the previous Decade Survey and strongly
supported in our report, is nearing completion.

SIRTF is NASA's next premier astrophysical observatory 
(concentrating on the infrared wavelengths) and has become
an important part of the agency-wide Origins program.  
The two Gemini telescopes, the infrared-optimized \hbox{8-m} northern 
telescope and
the southern \hbox{8-m} telescope, are funded through an international
collaboration with important leadership provided by the NSF.  At the
time of this writing, it seems likely that the millimeter array (the
MMA)  will be included in the president's budget (for 1998) for a
three year study leading to construction (provided no insuperable
obstacles are encountered).

I have indicated the current status of the large programs in Table~\ref{lgpro}.
The status of each of the two Gemini 8-m telescopes 
is denoted by a check mark to represent
the fact that these observatories are in an advanced stage of construction.
I have denoted by plus marks the status of  SIRTF and the MMA to indicate
that these major facilities are currently in the development stage.

\begin{table}[htb]
\centering
\tighten
\begin{minipage}[htb]{3.5in}
\caption[]{Large Programs\protect\label{lgpro}}
\begin{tabular}{lc}
\multicolumn{1}{c}{Program}&Status\\
\noalign{\smallskip\hrule\medskip}
Space Infrared Telescope Facility (SIRTF)&+\\
Infrared-optimized 8-m telescope&\check\\
Millimeter Array (MMA)&+\\
Southern 8-m telescope&\check\\
\end{tabular}
\end{minipage}
\end{table}

Our top three Moderate Programs, shown in Table~\ref{modpro},
 have all been successfully begun and the status of each one is 
indicated by a check mark.

Our highest priority, adaptive
optics, was largely declassified, and we now have access to important
developments initiated in connection with national security activities.
The civilian agencies have collaborated with the national security
agencies in  enabling this technology to be developed efficiently for
astronomical purposes.  We have a dedicated spacecraft for the FUSE
ultraviolet mission, and the contractors have been selected to build
 and fly the SOFIA airborne infrared  telescope facility. The  
high-resolution upgrade for the
Fly's Eye has been effectively achieved, and the marvelous extension to
the highest energy cosmic rays, the AUGER project (not a mature 
proposal at the time of the Decade Survey), seems almost
certain to occur.  I am personally very excited about the astronomy
and the physics that will be possible with AUGER.

The four intermediate priority Moderate Programs
listed in Table~\ref{modpro} have been pursued with moderate success.  For
example, the Astrometric Interferometry Mission (AIM) has been
included as part of an ambitious NASA initiative.  
Collaborations have been formed and further 
discussions are underway to establish
shared private 4-m and larger telescope consortia.
The two
Moderate Programs at the lower end of our priority listing, LEST and
the VLA extension, were casualties (very much regretted) of the
budget stringencies.

\begin{table}[htb]
\centering
\tighten
\begin{minipage}[htb]{4in}
\caption[]{Moderate Programs\protect\label{modpro}}
\begin{tabular}{lc}
\multicolumn{1}{c}{Program}&Status\\
\noalign{\smallskip\hrule\medskip}
Adaptive optics&\check\\
Dedicated spacecraft for FUSE&\check\\
SOFIA&\check\\
Delta-class Explorer acceleration\ \ (SMEX, MIDEX)&+\\
Optical and infrared interferometers&+\\
Several shared 4-m telescopes \ \ (private)&?\\
Astrometric Interferometry Mission (AIM)&+\\
High Resolution Fly's Eye&\check\\
Large Earth-based Solar Telescope (LEST)&0\\
VLA extension&0\\
\end{tabular}
\end{minipage}
\end{table}

I want to recount an illustrative anecdote about Table~\ref{modpro}.
In the months that followed the release of the Decade Survey, I gave
many talks about our recommendations and the exciting science that the
recommendations could make possible.  I always showed the full
versions from the report of Tables~\ref{lgpro}--\ref{smallpro}, which
included  estimates for the cost of each project.  I asked the
audience during each talk if they noticed anything unusual about
Table~\ref{modpro}.  Almost no one noticed the feature that I found
most revealing.  But, when I gave a talk to the faculty and trustees
of the Institute for Advanced Study, one emeritus professor
immediately noticed what I found remarkable.  George Kennan, former
Ambassador to the Soviet Union and a distinguished scholar of Russian
Studies, raised his hand and said:  ``Your highest priority
recommendation is by far the least expensive.  From all my years in
dealing with governments, I cannot remember another example of the
least expensive recommendation being the highest ranked.''

We prioritized according to scientific importance.

The committee recommended that an increased emphasis be given
in the astronomy research budget to small and moderate programs.
The committee did not prioritize small programs,
recognizing that the agencies could use peer review for small initiatives
to respond quickly to new scientific or technological developments.
However, all three of our Illustrative Small Programs, see Table~\ref{smallpro},
listed as high
priority are under development.

\begin{table}[htb]
\centering
\tighten
\begin{minipage}[htb]{3.3in}
\caption[]{Illustrative Small Programs\protect\label{smallpro}}
\begin{tabular}{lc}
\multicolumn{1}{c}{Program}&Status\\
\noalign{\smallskip\hrule\medskip}
Two-micron survey (2 $\mu$ SS)&\check\\
Infrared instruments&\check\\
Cosmic background explorer (MAP, CBI)&\check\\
\end{tabular}
\end{minipage}
\end{table}

The Microwave Anisotropy Probe (MAP) and the 
Cosmic Background Imager (CBI)
are two complementary instruments, one in space (MAP) and
 one ground-based (CBI), which,
together will observe anisotropy in the microwave background radiation
over angular scales from a few arc minutes to many degrees. I am 
proud  that the decade survey helped to facilitate these 
crucial experiments, which otherwise might have had  particular difficulty
in obtaining adequate funding.

These are the things that worked for us: enlisting as committee
members active research scientists eager to finish the job and get
back to their own work; recruiting an effective executive secretary;
insisting that the NRC provide adequate budgeting and staff support; having a logical
plan and a specific timetable for completing the report; listening to
everyone who wanted to be heard; concentrating on issues within
the committee's competence, in our case, scientific priorities; having
a talented editor (Susan Maurizi) who could sharpen the final report; and working
with a community that believes it is better for astronomers to make
imperfect judgments about priorities for astronomy than it is to
leave the decisions to Washington administrators.

In the years that have followed the publication of the Decade Survey,
I have made many visits to Washington to discuss specific projects
with staff and members of Congress, with people from OMB and OSTP, and
with senior leaders at NSF, NASA, and DOE.  I could get in to talk to
these important decision makers and could expect a sympathetic
reception because the Decade Report had a
favorable reputation for having set scientific priorities based upon a
consensus within the astronomy community.  These visits were, I believe,
particularly useful in helping to make possible the SIRTF, SOFIA, MMA,
and cosmic background explorer projects, and to facilitate the
declassification of prior DOD work, and the initiation of new research, on
adaptive optics.

\section{Personal Remarks}

I enjoyed very
much participating in the Decade Survey.
I did not expect to be able to say this when I started, but it is true.
I learned a great deal
about science and people from the process. 
Many individuals selflessly pulled together to make the survey a
success and I am grateful to each of them for the shared experience.

I was  lucky that Jerry Ostriker was also in Princeton.  
Jerry had unique
experience as a major participant in both the Greenstein and the Field
committee 
decade surveys; he also chaired the project initiation 
subcommittees that established
the Field committee and our survey.
I was not involved in the previous surveys nor with other NRC
committees, so it was immensely helpful to me that Jerry 
understood from the inside how the NRC and the
NAS worked. Jerry shared generously his organizational skills 
and his scientific 
insights. 

I am especially  grateful to Chas Beichman, who
served both as Executive Secretary and as a member of the survey
committee. Chas took leave from his normal research job 
to become a member of the Institute for Advanced Study during the
duration of the survey.  
The only argument I had with Frank Press during the entire activity of
the Decade Survey 
regarded Chas's presence in Princeton.  Frank's initial position was 
that, according to National Academy rules, the oversight of the
survey had to be maintained in Washington within the purview of the
National Research Council. I said if that was the case then I could
not serve, because I needed a ``second-in-command'' 
in Princeton in order to work
efficiently on the survey  while continuing my own research. We
compromised: Frank found a description  of  Chas's appointment that
allowed him to be in Princeton.  

Chas understood the big picture.  In addition, he knew exactly what things
had to get done and he made sure that they got done on time. When it 
was necessary to get the job completed, Chas did whatever background
research,  writing, or editing  was required.
I urge future survey
chairs to try to make sure that they have similarly strong support.

I am proud of what the Decade Survey accomplished. 
We all worked together and our scientific programs were
improved  because of the collaborations. 
Our report was welcomed and adopted
by the National Science Foundation and the National Aeronautics and
Space Administration.  It was widely praised in Congress and within the
Executive Branch.  The media coverage was large and favorable. As a
 community, we are frequently  held up as an example to other groups  of
what can be accomplished by forming a consensus about scientific
priorities. The process and recommendations 
of our survey were
sufficiently persuasive that we have  achieved a surprisingly high
fraction of what we proposed (see Tables~\ref{lgpro}--\ref{smallpro}).
And, very importantly for me, I
made a lot of close friends with whom I hope to 
share many pleasant  experiences in the future.

\end{document}